\pdfoutput=1
\documentclass{JINST}

\title{Characterizing Quantum-Dot-Doped Liquid Scintillator for Applications to Neutrino Detectors}

\author{Lindley Winslow$^a$\thanks{Corresponding
author.}~ and Raspberry Simpson$^a$\\
\llap{$^a$}Massachusetts Institute of Technology,\\
  77 Massachusetts Ave Cambridge, MA 02139, USA\\
  E-mail: \email{lwinslow@mit.edu}}

\abstract{Liquid scintillator detectors are widely used in modern neutrino studies. The unique optical properties of semiconducting nanocrystals, known as quantum dots, offer intriguing possibilities for improving standard liquid scintillator, especially when combined with new photo-detection technology. Quantum dots also provide a means to dope scintillator with candidate isotopes for neutrinoless double beta decay searches. In this work, the first studies of the scintillation properties of quantum-dot-doped liquid scintillator using both UV light and radioactive sources are presented.}


\begin{document}

\section{Introduction}

\begin{figure}
\begin{center}
\includegraphics[width=0.75\columnwidth]{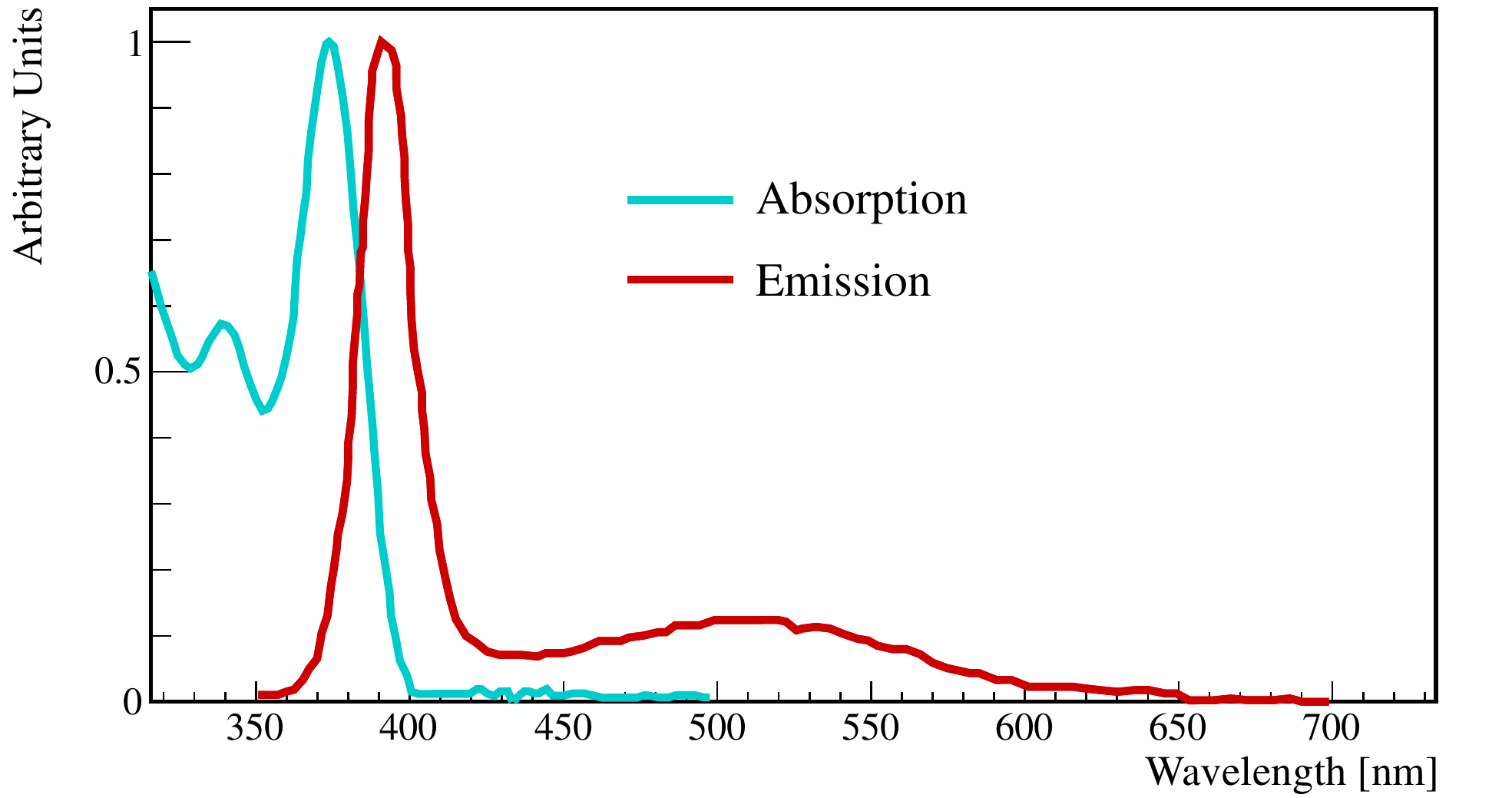} 
\end{center}
\caption{\label{diagram} Typical absorption and emission spectra for quantum dots. The units are arbitrary. Data taken from NN-Labs for 375~nm CdS dots\cite{NNLabs}.}
\end{figure}

Quantum dots are semiconducting nanocrystals. Due to their small size (2-10~nm), quantum confinement effects dominate, and the optical and electrical properties of the quantum dots are directly proportional to their size. As a result, the behavior of quantum dots bears a closer resemblance to that of single atoms than to bulk semiconductor, with smaller dots having a larger band gap. In fluorescence applications, this leads to smaller dots absorbing and re-emitting higher energy (shorter wavelength) photons. Due to the quantum nature of the process, the re-emission happens in a narrow resonance around the characteristic wavelength determined by the size of the dot. Example absorption and re-emission spectra are shown in Fig.~\ref{diagram}.  The synthesis of quantum dots permits precise control of dot size, allowing for fine tuning of the absorption and re-emission spectrum of the dots. 

An organic shell of ligands allows the quantum dots to be suspended in a variety of organic solvents and water, permitting diverse applications. 
Today the industrial uses of quantum dots range from tags for biological imaging to coatings for improved LED-lighting and solar cells.   While present commercial production yields only about 1~kg per year~\cite{Nanoco}, quality~\cite{QdotHighLum} and production techniques are rapidly advancing~\cite{QdotBusiness}, and one can expect the cost and commercial availability of quantum dots to improve considerably in the near future. On the other hand, the application of quantum dots to particle detection is not well-advanced.   Initial applications of quantum dot films for gamma ray~\cite{LetantScint}, low energy electron~\cite{CampbellScint}, and thermal neutron~\cite{WangZnS} detection have been explored; however, a more natural application for quantum dots may be as an additive to liquid scintillator. This is because certain synthesis methods suspend them in the organic solvent toluene -- an excellent scintillator. 

Liquid scintillators are the preferred technology for large-scale, multi-ton experiments due to the low cost of instrumentation.  Quantum-dot-doping for these  scintillator detectors may be appealing for several reasons. The most obvious is that this offers a perfectly tunable wavelength shifter that can be matched to new photon detection devices. Depending on the quantum efficiency of the dots, this could lead to a scintillator with a larger light-yield.  Also, as will be explored in the following section, the signal could be used to develop novel event identification. 

Beyond this, particle physics applications for quantum dots originate from the materials that form the dots themselves.
Quantum dots are typically made from binary alloys such as CdS, CdSe, CdTe, and ZnS.  A push to produce Cd-free quantum dots has led to the development of phosphor-based rare-earth dots~\cite{Nanostore}.  Therefore, quantum dots provide a method to dope scintillator with heavy metals and rare-earth elements, which is normally a difficult undertaking~\cite{DCScint}.  A Cd-based scintillator is particularly interesting for two reasons. Since $^{113}$Cd has the second highest thermal neutron capture cross-section with a gamma cascade totaling 9~MeV~\cite{CdSpec}, it is ideal for antineutrino measurements using inverse beta decay, $\bar{\nu}_e+p \rightarrow e^{+} + n$. Cd is also interesting because of two other isotopes.  $^{116}$Cd is  a double $\beta^{-}$ decay candidate, and $^{106}$Cd is one of only six isotopes capable of both double $\beta^{+}$ decay and double electron capture~\cite{Barabash}. It is interesting to note that Se, Te, and Zn all have double $\beta$ decay candidates as well.

\begin{figure}
\begin{center}
\includegraphics[trim = 0mm 10mm 0mm 5mm, clip, width=0.5\columnwidth]{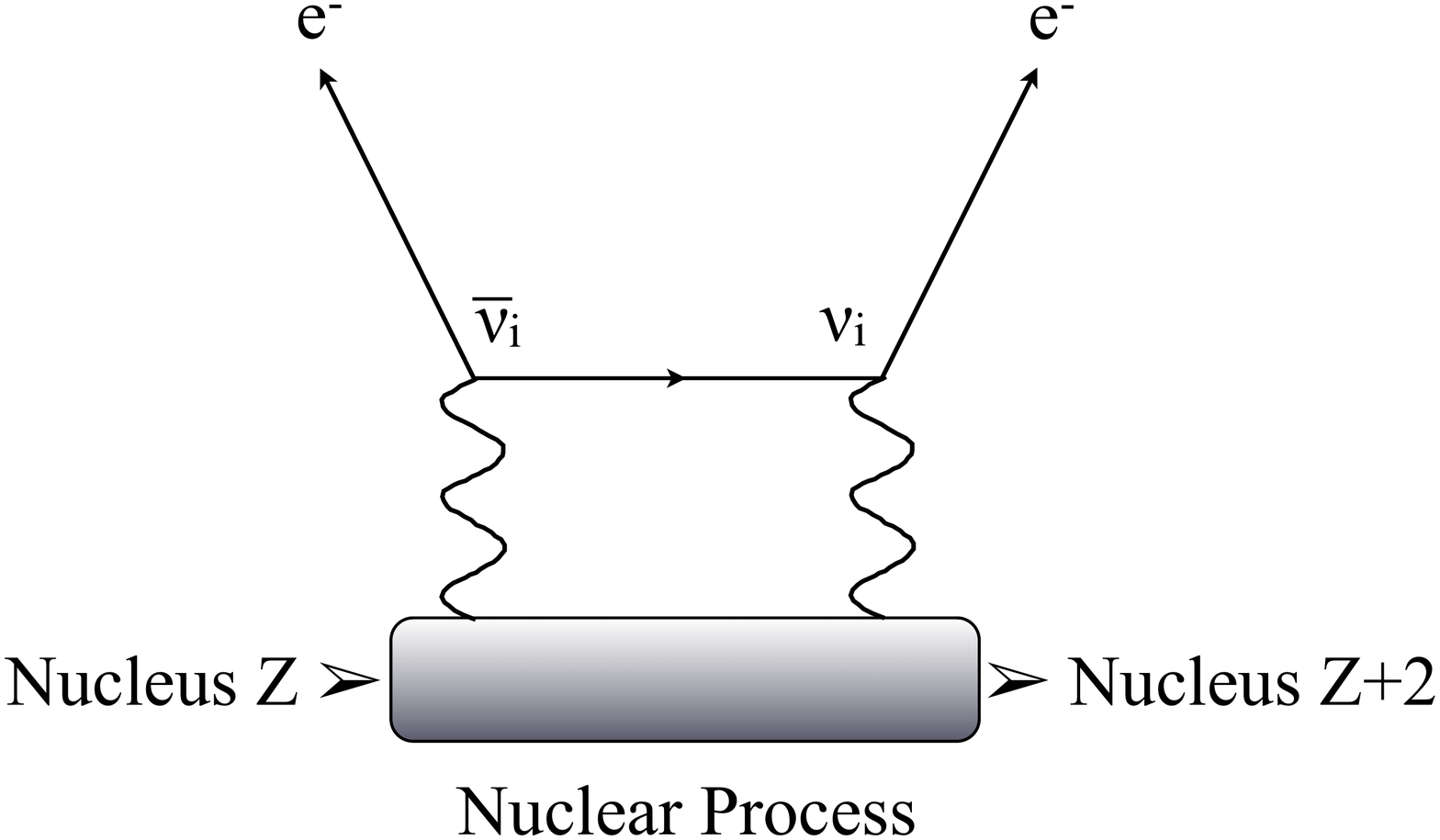} 
\includegraphics[width=0.4\columnwidth]{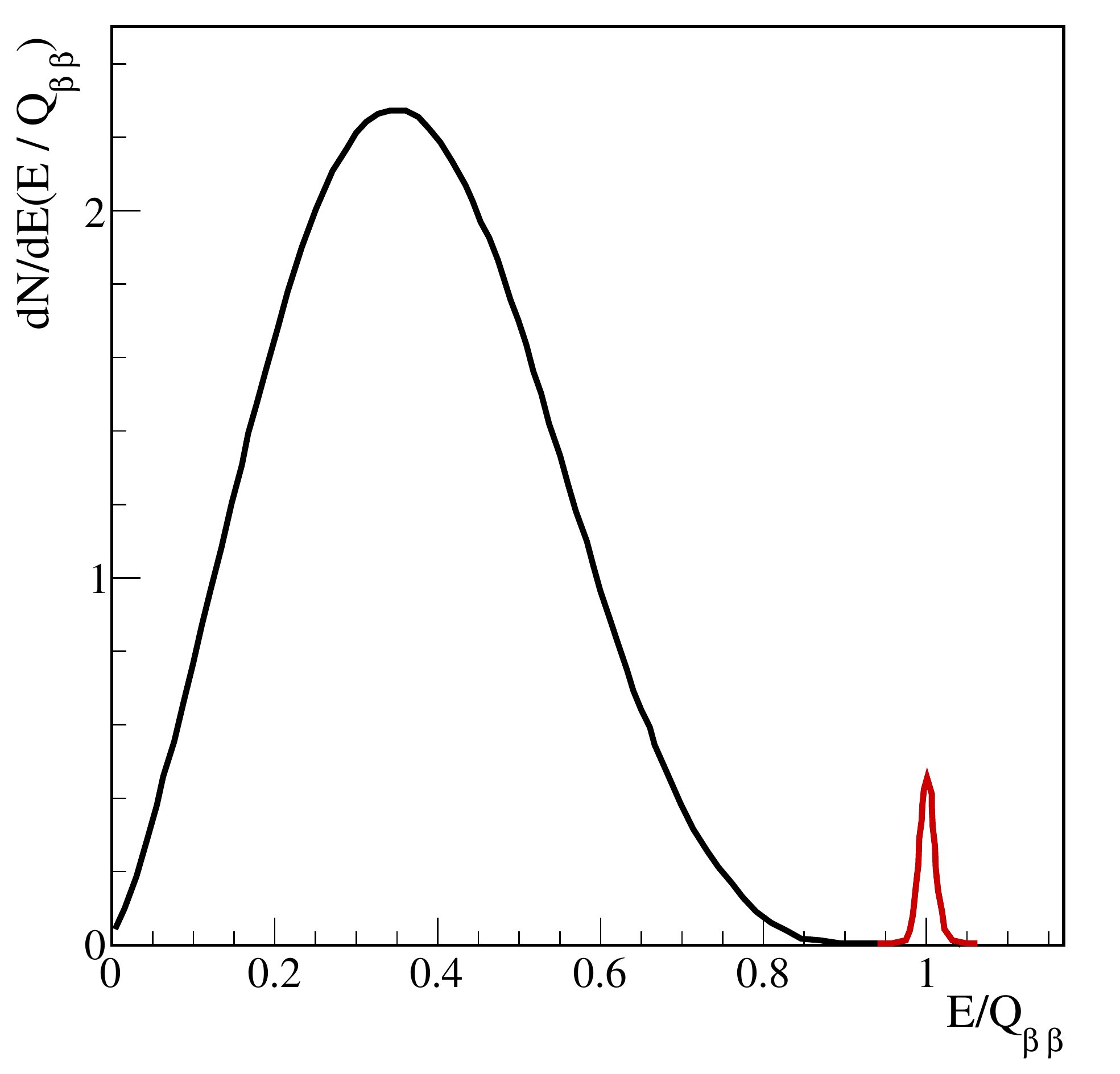}
\end{center}
\caption{\label{0nubb}  Left: Feynman diagram for neutrinoless double beta decay; Right: cartoon of the spectrum of electron energies from double beta decay. The red section at the endpoint $Q_{\beta \beta}$ indicates those from neutrinoless double beta decay, see also Ref.~\cite{RevBB}.}
\end{figure}

\section{Quantum-Dot-Doped Scintillator for Neutrinoless Double Beta Decay Studies}

Neutrinoless double beta decay is a beyond-Standard-Model process that tests for the Majorana nature of neutrinos. The decay can only occur in elements where the transition from atomic number $Z$ to $Z+2$ through two simultaneous $\beta$-decays is more energetically favorable than the single $\beta$-decay transition to $Z+1$.   As seen in Fig.~\ref{0nubb}~(left),  if neutrinos are their own antiparticle,  forming an internal propagator in the Feynman diagram, then two mono-energetic electrons will be produced. This signal sits at the endpoint of the two neutrino double beta decay spectrum, as illustrated in Fig.~\ref{0nubb}~(right). This cartoon is highly exaggerated. The neutrinoless double beta decay process will be very rare (if it exists at all) compared to the two neutrino double beta decay process. The latter has been observed in $^{116}$Cd~\cite{SoloRef, CobraRef}.

In rare-event searches, particle identification and event topology are key for reducing backgrounds. In searching for neutrinoless double beta decay, one would like to image the two electrons emerging and, if possible, reconstruct their energies~\cite{RevBB}.  This allows one to reject background events and to look for new physics in the electron angular correlations~\cite{0nuBBNewPhys}. In a scintillator-based detector, the emerging electrons with energies in the MeV range produce directional Cerenkov light as well as isotropic scintillation light. Some fraction of the Cerenkov light is produced at wavelengths too large to be absorbed and re-emitted by the scintillator.  The light above the absorption cutoff will propagate directly across the detector, retaining the electron directional information, and will not be slowed by scintillation processes.  With sufficient timing and photo-cathode coverage, one could use this Cerenkov light to image the electrons from neutrinoless double beta decay. 
\begin{figure}
\begin{center}
\includegraphics[width=0.75\columnwidth]{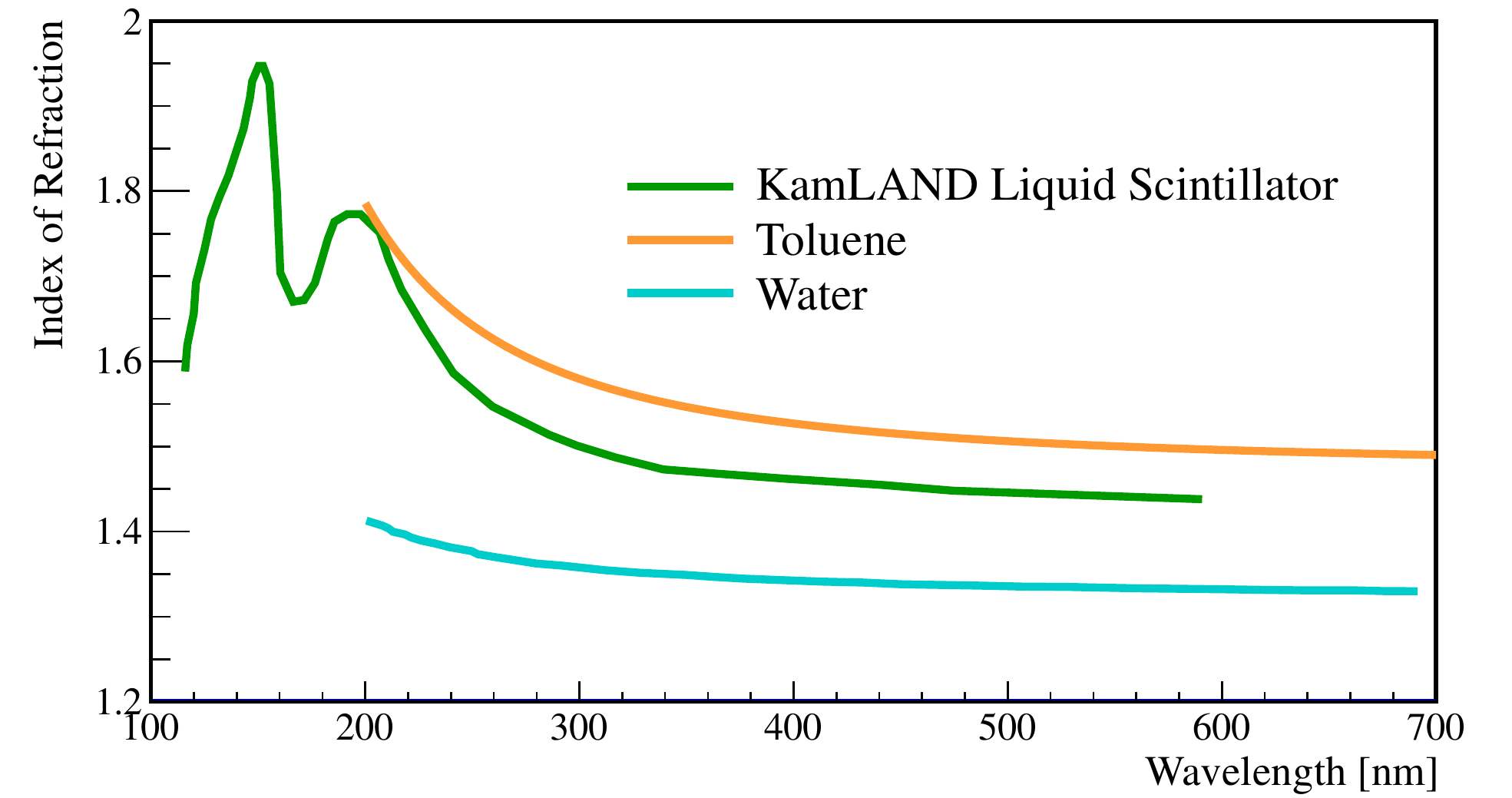} 
\end{center}
\caption{\label{indexOfRefraction}Index of refraction for a typical scintillator, KamLAND Scintillator~\cite{OlegThesis}. The index of refraction for plain toluene~\cite{TolIndex} and water~\cite{WaterCenkovRef} are shown for comparison. }
\end{figure}

The number of scintillation photons is much larger than the number of Cerenkov photons. The same quantum dots that provide the double beta decay isotope could be used to tune the scintillator's emission spectrum to shorter wavelengths and therefore enhance the Cerenkov signal relative to the scintillation signal. This assumes that the absorption cutoff is at even shorter wavelengths. The number of Cerenkov photons depends on the index of refraction of the detector material and the speed of the particle. The number of Cerenkov photons as a function of wavelength and distance is given by
\begin{equation}
\label{myEquation}
\frac{dN}{d\lambda dx} = \frac{2 \pi \alpha Z^2}{\lambda^2} ( 1 - \frac{1}{\beta^2 n(\lambda)^2} )
\end{equation}
where $n(\lambda)$ is the wavelength-dependent index of refraction.  The index of refraction of scintillators like that made for KamLAND~\cite{OlegThesis} or toluene~\cite{TolIndex} is higher than that of water~\cite{WaterCenkovRef} as summarized in Fig.~\ref{indexOfRefraction}.

\begin{figure}
\begin{center}
\includegraphics[width=0.75\columnwidth]{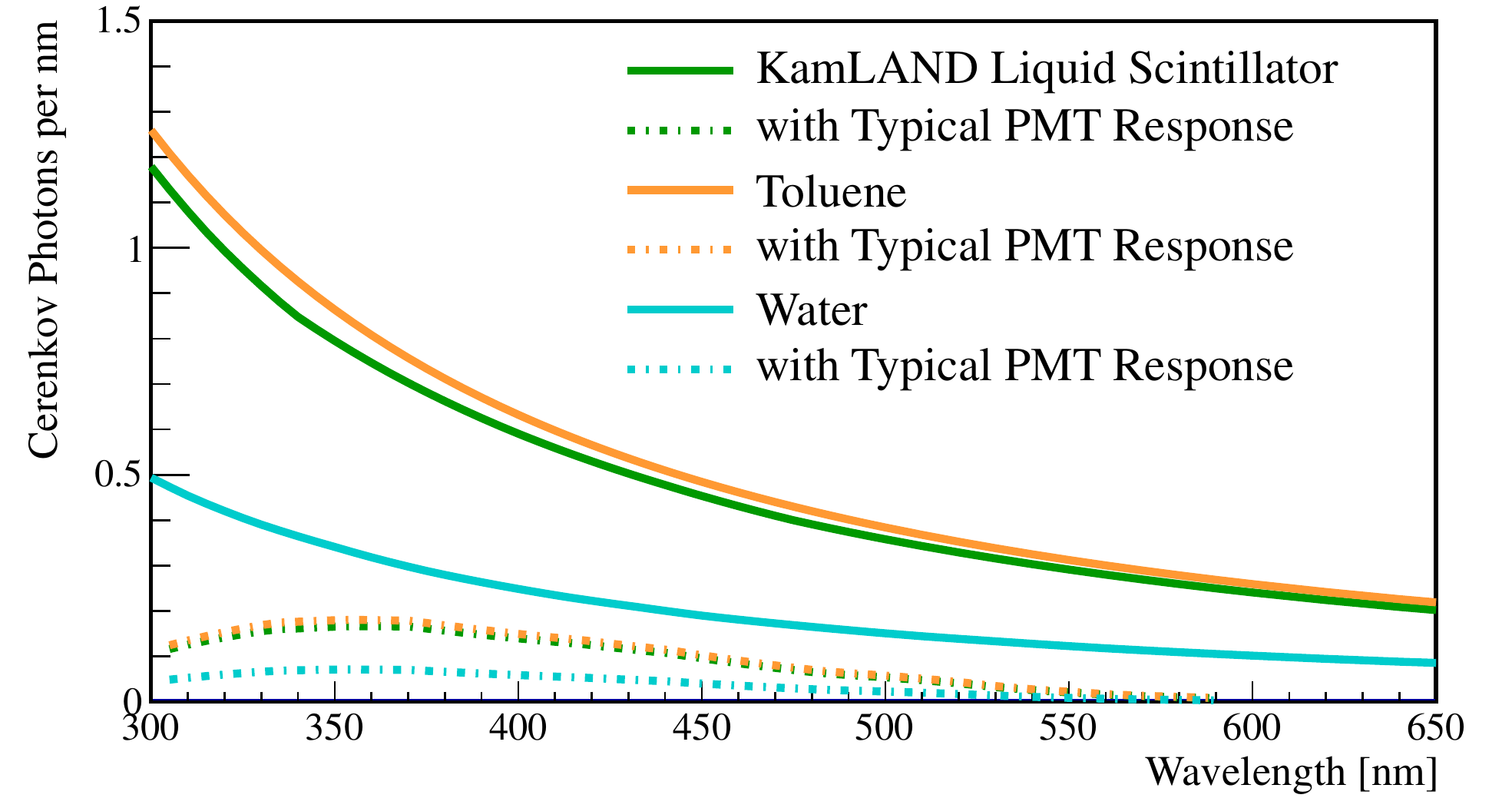} 
\end{center}
\caption{\label{Cerenkov} The spectrum of Cerenkov photons produced by a 1~MeV electron in several materials.}
\end{figure}

We can use Eq.~\ref{myEquation} to calculate the number of Cerenkov photons produced by a 1~MeV electron, a typical energy of one of the electrons from neutrinoless double beta decay. The results of this calculation are shown in Fig.~\ref{Cerenkov} with and without the response of a typical photomultiplier tube (PMT)~\cite{OlegThesis}. In Table~\ref{CTable}, we summarize the integrated photons between 400~nm and 550~nm as well as 360~nm to 550~nm.  This shows that if one were to use quantum dots to tune the absorption cutoff of a scintillator down to 360~nm, the number of Cerenkov photons would be increased by 40\%. 

In order to detect 15-20 photons from the Cerenkov light of a 1~MeV electrons, the photo-cathode coverage of the detector will need to be close to 100\%.  Also, the timing of the photo-detectors must be a fraction of a nanosecond to differentiate it from the scintillation signal.  These specifications match those of the Large Area Picosecond Photo Detectors (LAPPDs) that are now under-development~\cite{LAPPDSum, LAPPDTDR}.  The LAPPD collaboration is also working on photo-cathodes with improved quantum efficiency and photo-cathodes with high quantum efficiency in wavelengths above 400~nm. The latter could be used to enhance the Cerenkov signal at longer wavelengths. 

\begin{figure}
\begin{center}
\includegraphics[trim = 0mm 15mm 0mm 10mm, clip, width=0.75\columnwidth]{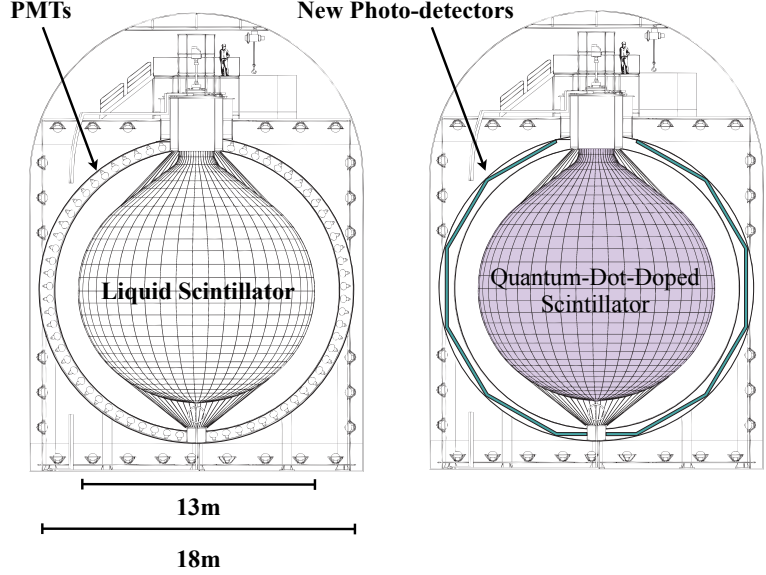}
\end{center}
\caption{\label{futureexp} The KamLAND~\cite{KamXen,KamFirst} experiment: the kiloton of plain liquid scintillator could be replaced with quantum-dot-doped scintillator to search for neutrinoless double beta decay.}
\end{figure}

The next generation of double beta decay experiments needs to instrument $\sim$1~ton of  isotope to explore the double beta decay parameter space corresponding to the inverted hierarchy for neutrino mass. An experiment aiming to explore the parameters space corresponding to the normal hierarchy will need to instrument  $\sim$10~tons of isotope. A quantum-dot-doped scintillator can be used in an experiment like KamLAND to search for neutrinoless double beta decay \cite{KamXen,KamFirst}. As is shown in Fig.~\ref{futureexp}, KamLAND's kiloton of plain liquid scintillator could be replaced with quantum-dot-doped scintillator. If photo-cathode coverage was increased with new photo-detection devices, the energy resolution would improve and more event topology information could be extracted; therefore, increasing the sensitivity to neutrinoless double beta decay.  In such an experiment, quantum dot concentrations of 1~g/L are needed to be sensitive to the inverted hierarchy and concentrations of 10~g/L are needed to address the normal hierarchy. These are typical concentrations for secondary wavelength shifters and for metal doping in reactor antineutrino experiments; however, it is critical to retain the total light yield and attenuation length of standard liquid scintillator.

\begin{table}
\caption{The calculated number of Cerenkov photons between 360-550~nm and 400-550~nm for some typical detector materials. The calculation with and without the PMT quantum efficiency as a function of wavelength is shown.}
\begin{tabular}{|c|c|c|c|c|}
\hline
 & \multicolumn{2}{|c|}{Number of Photons} & \multicolumn{2}{|c|}{with PMT Efficiency}\\
 \cline{2-5}
 & 400~nm & 360~nm & 400~nm & 360~nm\\\hline
Toluene & 65.8 & 94.0 & 12.0 & 18.5 \\ \hline
KamLAND Scintillator & 61.5 & 87.7 & 11.1 & 17.3\\ \hline
Water & 26.0 & 37.0 & 4.7 & 7.3 \\\hline
\end{tabular}
\label{CTable}
\end{table}

In order to explore the feasibility of this double beta decay application, we have studied quantum-dot-doped scintillators.
This initial suite of work is performed with 20~mL samples in scintillation vials. We use toluene with 5~g/L PPO (2, 5-Diphenyloxazole) as our standard scintillator. In the following studies, we add the toluene-suspended quantum dots to the toluene-based scintillator as a secondary wavelength shifter. The quantum dot concentration used for all samples is 1.25~g/L. We have chosen to use CdS core quantum dots because they have characteristic wavelengths in the 360-460~nm range where PMTs are most sensitive. CdSe dots typically have higher quantum efficiencies, but their typical wavelengths are in the region where the PMT quantum efficiency is dropping quickly. The quantum dots are purchased from two companies NN-Labs~\cite{NNLabs} and Sigma-Aldrich made by Nanoco~\cite{Nanoco}.

For this set of experiments we took no special care to avoid oxygen exposure, though a reduced-oxygen environment should improve the light output of scintillator and increase the life time of the quantum dots. Because of the small volume, we are not going to address the issue of attenuation length.  Large detectors like those shown in Fig.~\ref{futureexp} require attenuations lengths greater than 10~m. The overlap between the absorption and reemission spectra shown in Fig.~\ref{diagram} indicates that if the quantum efficiency of the dots is not sufficiently high, the attenuation length would be reduced. Both of these issues should be addressed in future work.

\section{Scintillator Studies:   Comparative Response to UV Light}

The first set of measurements is performed to compare the response of the quantum-dot-doped scintillator as a function of wavelength to the standard toluene and PPO scintillator. The sample vial is placed in a light-tight canister where it is excited using a 280~nm LED and readout using a StellarNet UV-VIS spectrometer. The sample spectrum is obtained from a 5-minute data sample taken in 30~ms exposures. A background spectrum with the LED off is acquired  for each sample and subtracted from the sample spectrum.

\subsection{Quantum Dot Selection}

\begin{figure}
\begin{center}
\includegraphics[width=0.75\columnwidth]{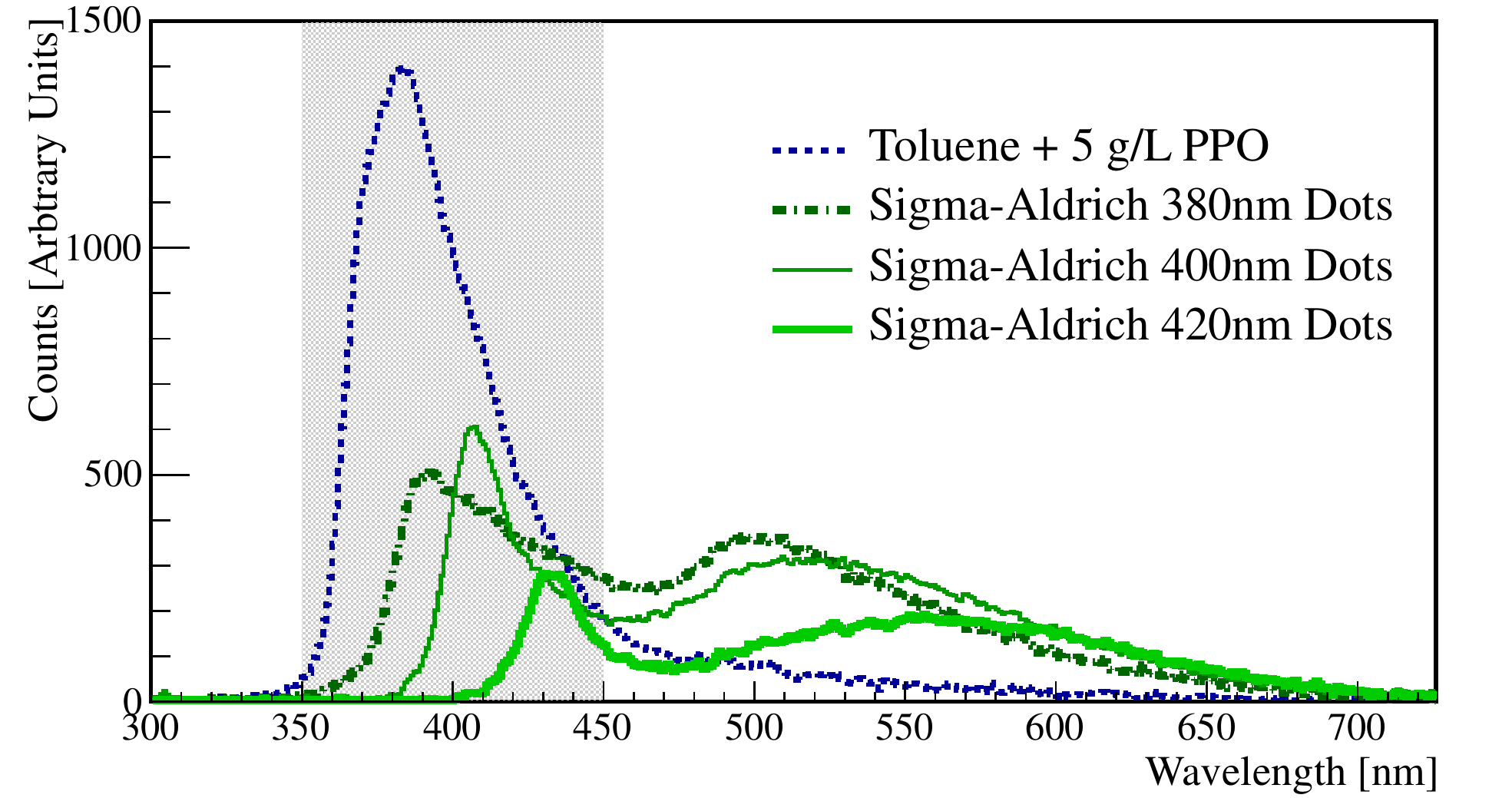} 
\includegraphics[width=0.75\columnwidth]{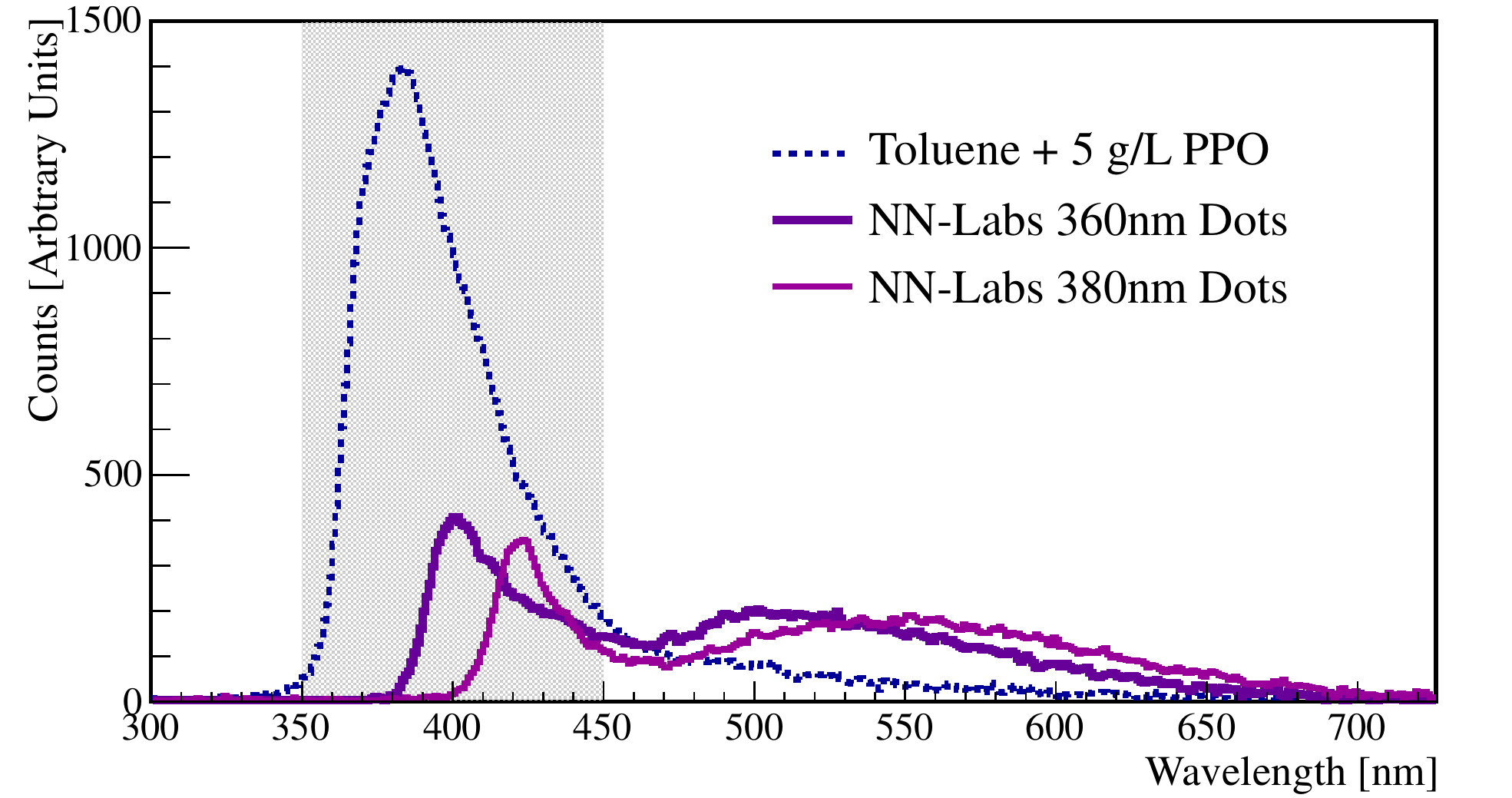} 
\end{center}
\caption{\label{uvsize} The response of different-sized quantum dot scintillator mixtures excited by 280~nm
  light.  The quantum dots are suspended in the 5~g/L PPO and toluene solution. The gray box indicates the peak sensitivity range of typical PMTs. }
\end{figure}

The emission spectra for the standard scintillator and the scintillator samples doped with quantum dots from Sigma-Aldrich and NN-Labs are shown in Fig.~\ref{uvsize}.  These data show that the quantum-dot-doped scintillator has lower light output than the standard scintillator. They also show that the emission spectrum of the scintillator is dominated by the  quantum dots. In fact, the longer wavelength photons from trap states on the surface of the dots are seen in all the dots at around 500~nm.  This indicates that quantum dots can be used to tune the spectral response of a liquid scintillator. 

The lower light output of the quantum-dot-doped scintillator is most likely due to the quantum efficiency of the dots. Improving quantum efficiency is a major area of research. 
The quantum dots purchased from Sigma-Aldrich have better light output than those from NN-Labs. In general, the light output of the smaller/shorter wavelength quantum dots from the same company are better than the longer wavelength dots, although the shortest wavelength dots show increased widths of their emission peaks.  We note the naming scheme for quantum dots is not consistent from company to company. For instance, the 360~nm NN Lab quantum dots are comparable in emission wavelength to the 400~nm Sigma Aldrich dots.

\subsection{Quantum Dot Aging and Batch Differences}

\begin{figure}
\begin{center}
\includegraphics[width=0.75\columnwidth]{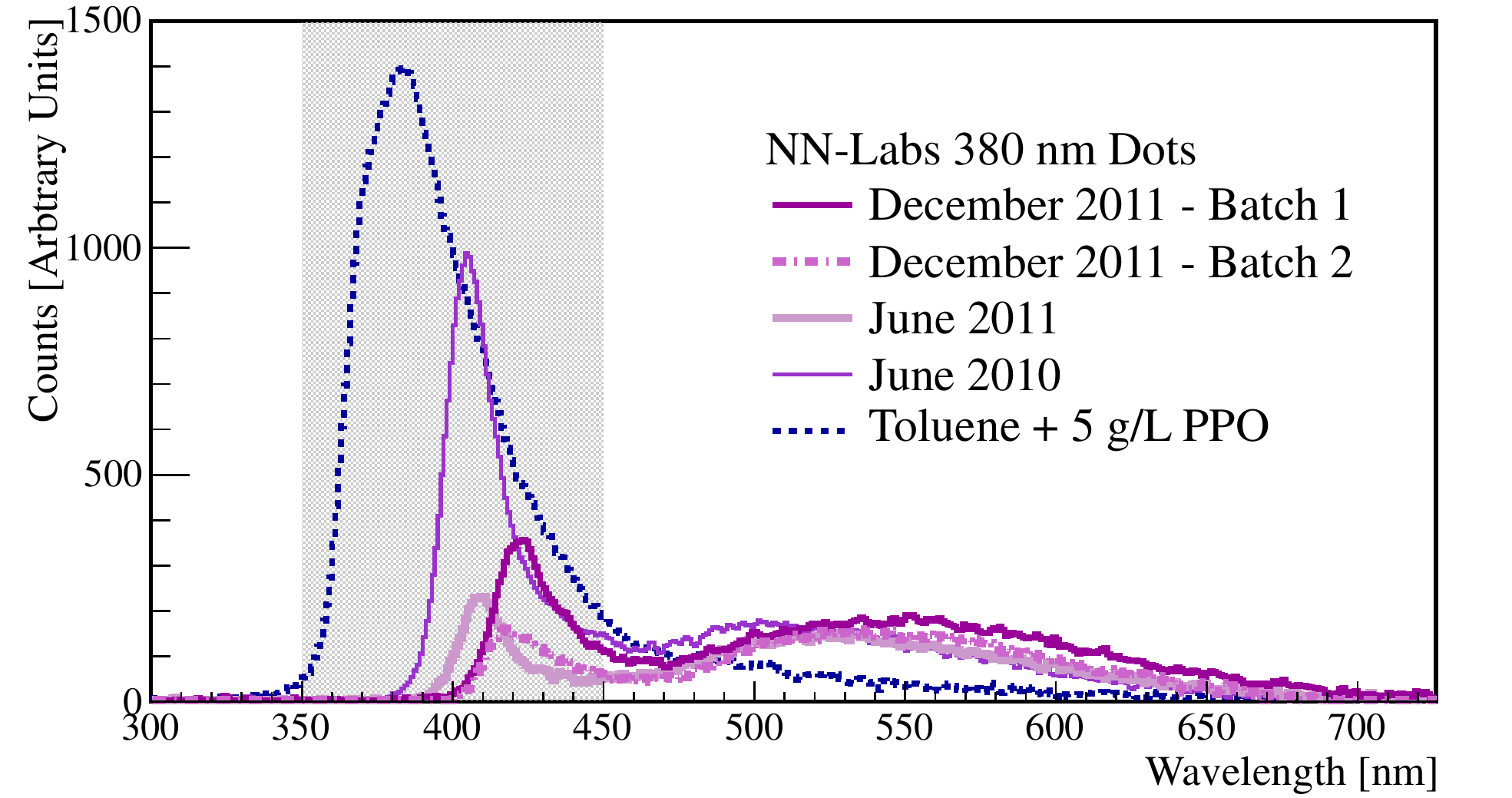} 
\end{center}
\caption{\label{uvbatch} The response of different batches NN-Labs 380~nm quantum dot scintillator mixtures excited by 280~nm
  light.  The quantum dots are suspended in the 5~g/L PPO and toluene solution. The gray box indicates the peak sensitivity range of typical PMTs. }
\end{figure}

A concern when constructing large liquid scintillator detectors is the degradation of the scintillator with age.  For quantum dots, it is known that the organic molecules that allow the dots to be suspended in organic solvents or water can also lead to the clumping of dots and the degradation of their performance. We purchased our first sample of quantum dots from NN-Labs in June of 2010.  In Fig.~\ref{uvbatch} we compare the samples purchased in June of 2010,  June of 2011, and two samples purchased in December of 2011 labelled Batch 1 and Batch 2.  We find no evidence at this time for the aging of the quantum dot scintillator. Our biggest issue is the batch-to-batch variation in both the characteristic wavelength and the quantum efficiency.  Sigma-Aldrich only recently started selling CdS quantum dots, so we do not have data on their batch-to-batch performance at this time.

\section{Scintillator Studies: Comparative Response to $^{90}$Sr Source}

The second set of measurements is performed to compare the response of quantum-dot-doped liquid scintillator relative to the standard toluene and PPO scintillator, and to ensure that the quantum dots do not interfere with the primary scintillation process. These studies are performed using a $^{90}$Sr pin source suspended in the sample. The decay of $^{90}$Sr goes to $^{90}$Y with a 0.5~MeV $\beta$. $^{90}$Y then decays with a 64-hour half-life emitting a 2.28~MeV $\beta$.  The response of the scintillator to MeV $\beta$'s is particularly relevant for neutrinoless double beta decay applications.

The scintillation light is collected by two Hamamatsu PMTs R1828-01 with 1.3~ns rise time and 0.55~ns transit time spread arranged perpendicularly to each other with the 20~mL sample at the intersection point.  For light-yield measurements both PMTs, PMT A and B,  are located 6~cm from the sample.  For timing measurements, one PMT,  PMT B is moved to 42~cm from the sample. The signals from the PMTs are digitized using the AlazarTech ATS9870 with 1~ns resolution and 100~mV range.  The signal in PMT A is delayed by 255~ns to avoid cross-talk during digitization. This PMT is also used as the trigger with the threshold set at 14.8~mV. The digitized waveforms are then analyzed using pulse-finding algorithms. The pulse charge is defined as the area under the pulse, and the pulse time is defined as the time of the peak of the pulse.

\subsection{Light Yield}

\begin{figure}
\begin{center}
\includegraphics[width=0.75\columnwidth]{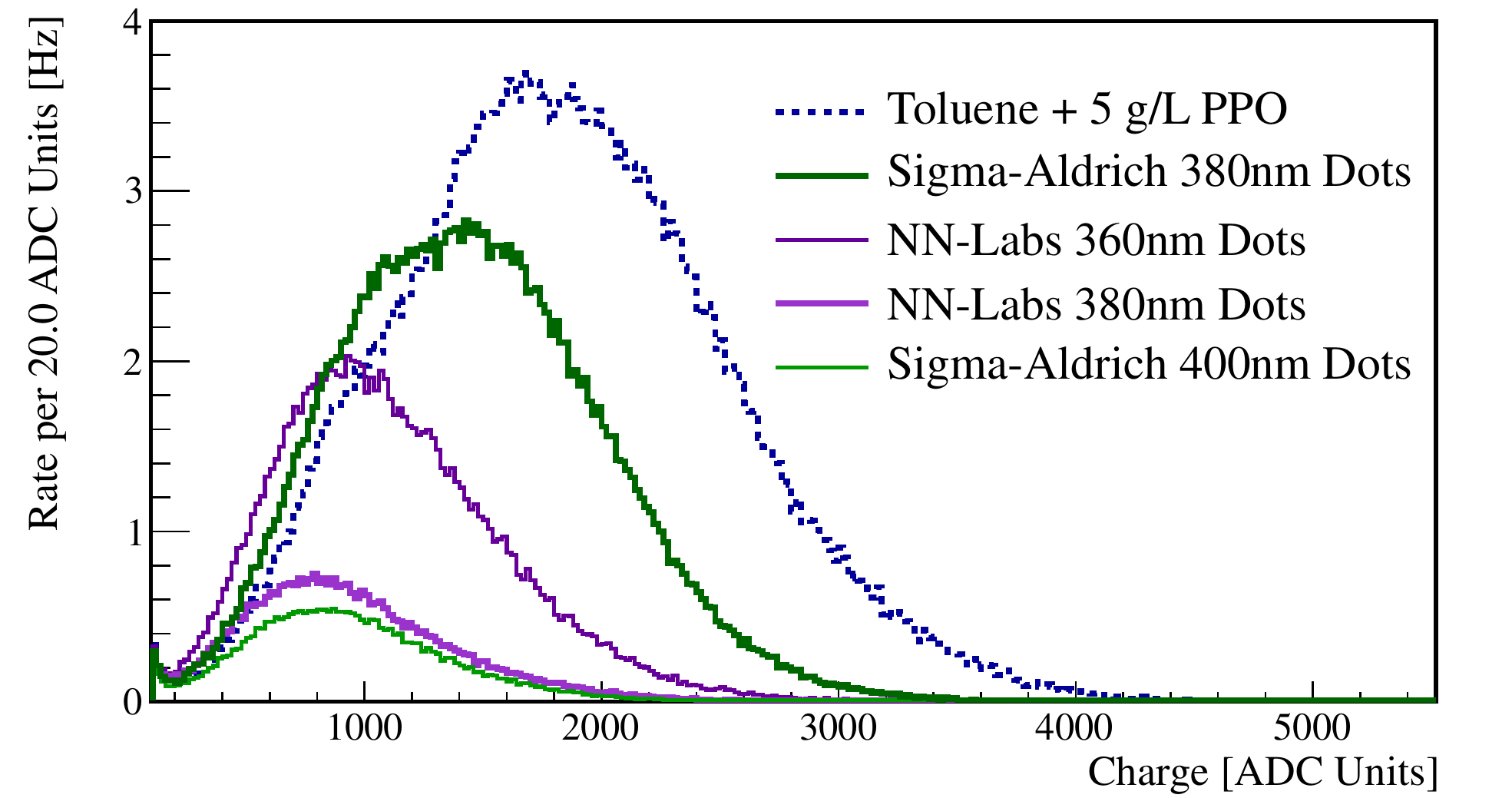} 
\end{center}
\caption{\label{lightyield} The total light-yield of different scintillator mixtures obtained with a $^{90}$Sr source. The NN-Labs 380~nm dots are the sample from June 2010.}
\end{figure}

The light-yield is inferred from the sum of the pulse charge collected in PMT B in a particular event.  Fig.~\ref{lightyield} summarizes the light-yield measurements for the standard scintillator and several quantum-dot-doped scintillators. The light-yield measurements show reduced light output from the quantum-dot-doped scintillator  compared to the standard scintillator. This is expected given the results of the UV response studies.  The increased light-yield from shorter wavelength dots also carries over from the UV response studies, however it is not clear why the 360~nm NN-Labs dots perform better than the 400~nm Sigma-Aldrich dots.  This last point seems to contradict the UV data and needs more investigation.

\subsection{Time Distributions}
\begin{figure}
\begin{center}
\includegraphics[width=0.75\columnwidth]{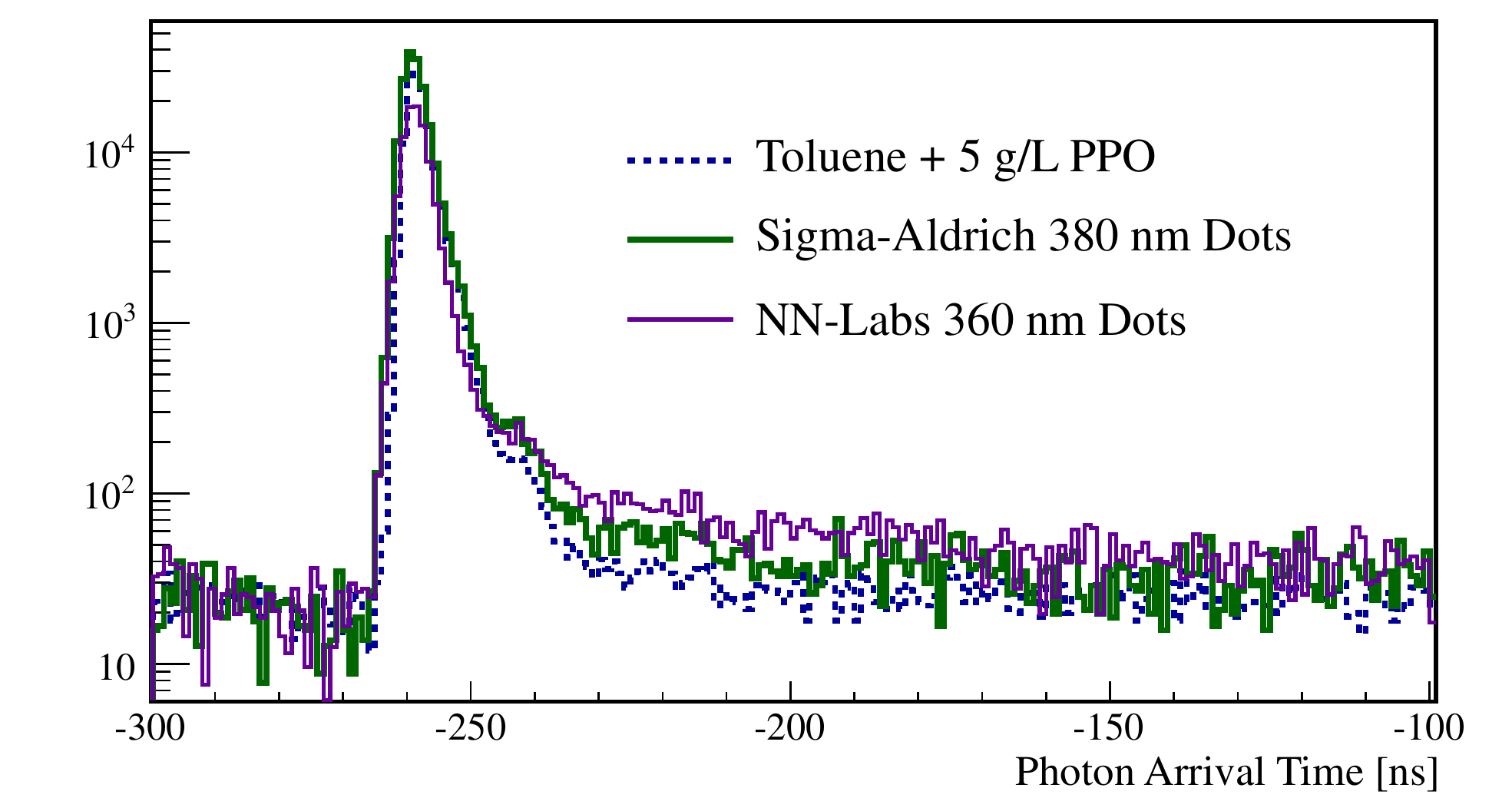} 
\end{center}
\caption{\label{timing} The timing response of different scintillator mixtures obtained with a $^{90}$Sr source. The signal is delayed by 255~ns. The background level of the samples is normalized to the background level of the Toluene + 5 g/L PPO at - 300~ns  because of different background levels in the runs and to allow for easier comparison.}
\end{figure}

The timing distribution of the liquid scintillators is determined by taking the difference between the trigger pulse in PMT A with all pulses in PMT B.  The signals in PMT A are delayed by 255~ns, so the distribution reflects this offset. The timing distributions for the mixtures with the largest light-yields are summarized in Fig.~\ref{timing}.  The distributions of the quantum-dot-doped scintillators are similar to the standard toluene and PPO scintillator.  The background levels or dark rate levels are higher in the quantum-dot-doped samples. For this reason, the time distributions are normalized to the background level in the toluene + 5~g/L PPO sample. This increased background level may be due to the known process of quantum dot ``blinking", but more study is needed.

To quantify the features observed in the timing distributions, they are fit to a model of scintillator time response using three exponential components\cite{ChristophThesis}. This model includes the response of the PMT. The results of the fit are presented in Table~\ref{TimeTable}. It is critical for the shortest and dominant component to be on the order of 2~ns to enable position reconstruction in a kiloton-scale detector. Note that this rise time is still large on the scale of the picosecond timing of LAPPDs. The quantum-dot-doped scintillators preserve the speed of the short component. The longest component is longer in the quantum-dot-doped scintillators.  This is interesting as the ratio of the short to the long component in some scintillators allows for the differentiation of more highly ionizing particles like alphas from electrons and gammas.  This is a feature that warrants closer study in the future.

\begin{table}
\caption{The results of fitting the measured time distributions to a model of the scintillator time response with three time components $\tau_i$.  The weight of each component is $q_i$ and the PMT response is included in the model. Toluene with 5~g/L PPO is abbreviated as Tol $+$ PPO. Sigma-Aldrich 380~nm dots are abbreviated as SA 380~nm. NN-Labs 360~nm is abbreviated NN 360~nm.}
\begin{tabular}{|c|c|c|c|c|c|c|}
\hline
Sample & $q_1$ & $\tau_1$ [ns] & $q_2$  & $\tau_2$ [ns]  & $q_3$ & $\tau_3$ [ns]  \\
\hline
Tol $+$ PPO & 0.94$\pm$0.01 & 1.73$\pm$0.03& 0.08$\pm$0.01 & 5.7$\pm$0.5& 0.004$\pm$0.001 & 45.9$\pm$23.4\\
\hline
SA 380~nm & 1.10$\pm$0.01 & 1.84$\pm$0.02& 0.09$\pm$0.01 & 6.5$\pm$0.4& 0.022$\pm$0.001 & 96.5$\pm$10.7 \\
\hline
NN 360~nm & 0.80$\pm$0.01 & 1.55$\pm$0.03& 0.06$\pm$0.01 & 10.9$\pm$0.7& 0.092$\pm$0.003 & 174.5$\pm$14.9\\
\hline
\end{tabular}
\label{TimeTable}
\end{table}

\section{Conclusions}

Quantum-dot-doped scintillators hold promise for future neutrino detectors, especially those searching for neutrinoless double beta decay.  Their properties when combined with new photo-detection technology appear particularly promising. This first round of experiments indicate that the quantum dots degrade the total light output of the scintillator, but not beyond the point where it would be applicable for the next generation of neutrinoless double beta decay experiments that are looking to instrument 1~kiloton of scintillator with $\sim$1~g/L of isotope. Also, the efficiency of quantum dots continues to improve, thus the scintillator merits continued study.  

As a next step, 1~L of scintillator will be instrumented to explore further the properties of quantum-dot-doped liquid scintillator.  This larger sample size will also make possible studies of attenuation length, a critical measurement for large-scale detectors. This 1~L detector may be a precursor to a larger 1~m$^{3}$ detector.  The goal of this larger detector would be to demonstrate the feasibility of Cerenkov imaging via a measurement of the two neutrino double beta decay in $^{116}$Cd. Approximately 1000 events are expected in a six-month run.

\acknowledgments
The authors thank the L'Or\'{e}al for Women in Science Fellowship for the support of equipment, and the National Science Foundation for support of personnel. We also thank E. Adanu and R. Jerry for their early work on the topic and the MSRP programs for their and R. Simpson's support. The authors thank Tess Smidt for editing the KamLAND drawings. Finally, the authors thank J.M. Conrad for useful discussions and the labspace to pursue this research.

\bibliographystyle{JHEP}
\bibliography{QDotPaper}
\end{document}